\documentclass{aastex}
\usepackage{spr-astr-addons}
\usepackage{url}

\RequirePackage{color}

\begin{document}

\title{New Results on the Ages of Star Clusters in Region B of M82}\shorttitle{Ages of Stellar Clusters in M82-B}
\shortauthors{Konstantopoulos et al.}

\author{I. S. Konstantopoulos\altaffilmark{1}}
\author{N. Bastian\altaffilmark{1}}
\author{L. J. Smith\altaffilmark{2,1}}
\author{G. Trancho\altaffilmark{3,4}}
\author{M. S. Westmoquette\altaffilmark{1}}
\author{J. S. Gallagher III\altaffilmark{5}}

\altaffiltext{1}{Department of Physics and Astronomy, University College London, Gower Street, London, WC1E 6BT, UK}
\altaffiltext{2}{Space Telescope Science Institute and European Space Agency, 3700 San Martin Drive, Baltimore, MD 21218, USA}
\altaffiltext{3}{Universidad de La Laguna, Avenida Astr\'{o}fisico Francisco S\'{a}nchez s/n, 38206, La Laguna, Tenerife, Canary Islands, Spain}
\altaffiltext{4}{Gemini Observatory, 670 N. A'ahoku Place, Hilo, HI 96720, USA}
\altaffiltext{5}{Department of Astronomy, University of Wisconsin-Madison, 5534 Sterling, 475 North Charter Street, Madison, WI 53706, USA}

\begin{abstract}
The post-starburst region B in M82 and its massive star cluster component have been the focus of multiple studies, with reports that there is a large population of coeval clusters of age $\sim$1~Gyr, which were created with a Gaussian initial mass distribution. This is in disagreement with other studies of young star clusters, which invariably find a featureless power-law mass distribution. Here, we present Gemini-North optical spectra of seven star clusters in M82-B and show that their ages are all between 10 and 300~Myr (a factor of 3-100 younger than previous photometric results) and that their extinctions range between near-zero and~4~mag~($A_V$). Using new {\it HST ACS-HRC U}-band observations we age date an additional $\sim$30 clusters whose ages/extinctions agree well with those determined from spectroscopy. Completeness tests show that the reported `turn-over' in the luminosity/mass distributions is most likely an artefact, due to the resolved nature of the clusters. We also show that the radial velocities of the clusters are inconsistent with them belonging to a bound region.

\end{abstract}

\section{Introduction}
Region B, in the local starburst galaxy M82, has been the object 
of debate in recent years, due to claims that it 
hosts a 1~Gyr-old, independent starburst region 
\citep{RdG01}. Furthermore, a turn-over 
was reported for the cluster mass distribution. 
This would make it unique among young cluster 
populations, which are generally accepted (and theoretically expected) to have power-law distributions.

As part of a larger project\footnote{We refer the reader to \citet[][multi-band photometry]{m82phot07} and \citet[][spectroscopy]{isk08a} for full details of source selection, data acquisition, reduction and analysis, and full spectroscopic and photometric results.} to study the cluster 
population of the entire galaxy, we have utilised 
multi-band ({\it UBVI}) {\it HST-ACS} photometry of 35 
Young Massive Clusters (YMCs) in the region and 
additionally, Gemini-North multi-object spectroscopy (GMOS-N)
of seven of them, from which we 
constrain extinctions, ages and radial velocities. We present these results in sections \S~2, 3 and 4 and a summary and conclusions in \S~5.

\section{Estimating the extinction}
The clusters in region~B have been presented in 
the past as having very low extinctions \citep{RdG03a}. 
Coupled with the claim that the 
region has a very high gas/dust extinction gradient, it was 
suggested that only the outermost layers of the 
region are visible. 

\begin{figure}[tbhp]
\begin{center}
\includegraphics[width=0.48\textwidth]{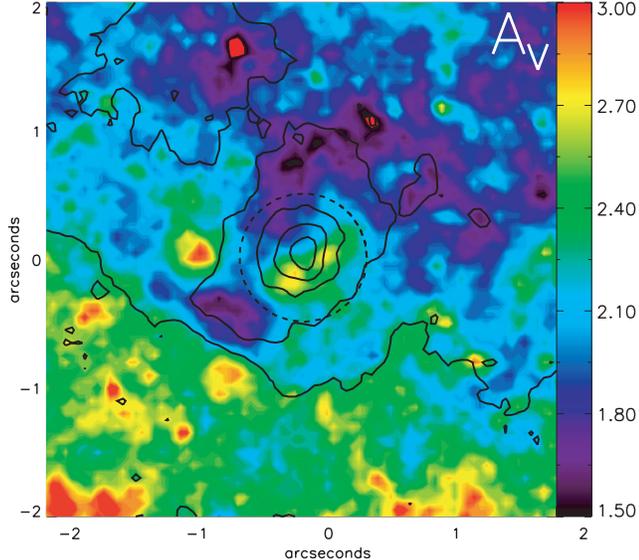}
\caption{Extinction map for cluster 91 (M82-H). The 
extinction is calculated for each pixel using the available 
{\it BVI} photometry (given an accurate spectroscopic age of 190~Myr). 
The solid lines represent isophotes in the F435W-band 
({\it V}) and the dotted circle denotes the aperture used for 
the cluster photometry. The colour map gives the 
extinction in magnitudes. The spatial scale of the map is
30$\times$30~pc. 
Note the strong dust lane running across the cluster, 
strongly influencing the photometry and photometrically 
derived values, such as age and extinction. 
}
\label{plot:ext}
\end{center}
\end{figure}

Using the three-dimensional maximum likelihood fitting photometric technique \citep[3DEF,][]{bik03}, 
and reddening 
maps in some cases, we find that the 35 clusters in our photometric 
sample have extinctions between near zero 
and $\sim$2.5~mags (in $A_V$, after foreground extinction correction). We also find pronounced effects of differential 
extinction across the face of some clusters (see 
Figure.~\ref{plot:ext}). While this leaves the spectroscopically measured properties unaffected, it complicates the measurement of cluster colours, and consequently, the photometric age determination.

The measured range of extinctions establishes region~B as one of the overall least 
obscured parts of the M82 disk, however, it does not necessarily imply that all visible clusters are on the surface. Our data
offer the interpretation that region~B presents a view deep into the body of the galaxy through a series of low extinction 
`portholes', similar to ``Baade's windows'' in the Milky Way. As a 
result of this, a greater number of 
clusters can be seen in region~B compared to the rest of the 
disk. This large number of clusters should not be treated as an intrinsic property for this part of the galaxy (such as evidence of a higher SFR in the region), but as an apparent one. In fact, {\it HST-NICMOS} near infrared observations of the galaxy \citep{alonso01,alonso03} show similar luminosities and cluster densities in both galaxy lobes.

\section{Age determination}
We use a number of independent photometric 
and spectroscopic methods in order to obtain 
the best possible age estimate for each 
cluster. All of these methods are based on the 
use of models and the goodness of fit between 
observation and model. 

\begin{figure}[bth]
\begin{center}
\includegraphics[width=0.48\textwidth]{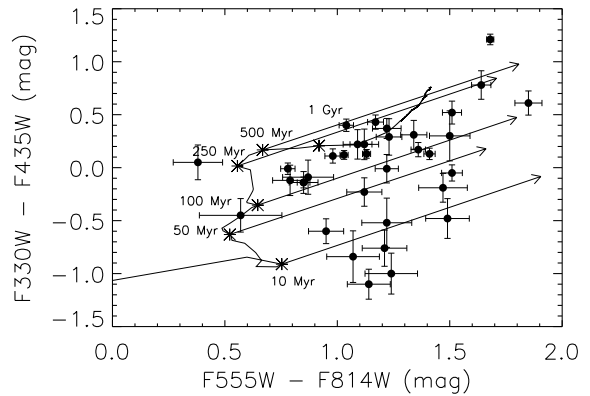}
\caption{$U-B$ vs $V-I$ plot for all 35 {\it U}-band selected clusters (this includes six of seven clusters for which we have spectra). The extinction vector has length $A_V=3$~mag. The cluster that falls outside the evolutionary track (cluster M82-H, presented in Figure.~\ref{plot:ext}) has an underestimated $U-$band flux. 
}
\label{plot:color}
\end{center}
\end{figure}

As a first estimate, we determine the ages from 
the comparison of {\it UBVI} magnitudes to cluster 
evolutionary synthesis models (GALEV models, Anders et al. 
2005), using the 3DEF method (demonstrated in Figure.~\ref{plot:color}). We then proceed to test 
these against two independent spectroscopic 
methods. 

\begin{figure}[bthp]
\begin{center}
\includegraphics[width=0.48\textwidth]{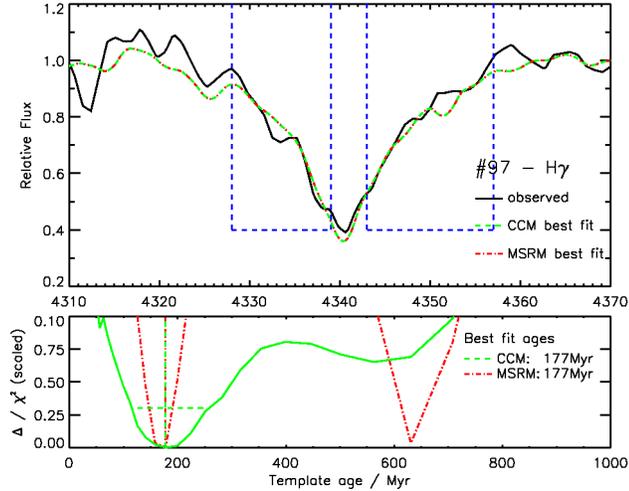}
\caption{The spectroscopic age determination routine for cluster 97 (H$\gamma$ line). 
The bottom panel shows the cumulative $\chi^2$ method (CCM) and model-spectrum residual method (MSRM) 
statistics in green and red respectively as a function of model 
age, with the best fits indicated by the vertical lines. The 
horizontal dashed line indicates the adopted error margin of twice 
the minimum $\chi^2$ value. The top panel shows the observed 
spectrum with the best fit model over-plotted.
}
\label{plot:age-fit}
\end{center}
\end{figure}

The more important of these methods is $\chi^2$
fitting of lower Balmer series lines (H$\beta$ and H$\gamma$) with 
models, specifically designed for young stellar 
clusters \citep{gd05}. We refer to this method as 
the cumulative $\chi^2$ method (CCM) and it is 
demonstrated in Figure.~\ref{plot:age-fit} for a randomly selected 
cluster. Its advantage is that it fits on the overall 
profile shape, rather than just on the line strength 
(like the simpler employed model-spectrum residual method, or MSRM). In all, we find 
spectroscopic and photometric methods to be in accord within the calculated 
errors. 

Overall, we find the age distribution of YMCs in 
region B to peak at $\sim$150~Myr, which is much 
younger than has been claimed in the past \citep[1~Gyr 
peak in][]{RdG01,RdG03a}. In addition, 
we find several of the clusters to have ages 
similar to M82-F \citep[measured as 60~Myr by][]{SnG}, which resides on 
the opposite side of the galaxy. 
In \citet{isk08b} we perform a similar analysis of the entire galaxy disk population and find the same characteristics for the age distribution of the galaxy. This coincidence of SF peak and spread in measured ages between region~B and the full galaxy sample reinforces our suggestion that this region is not `special', 
but is simply located behind a line of sight of 
relatively low extinction.

\section{Radial Velocities}
\begin{figure}[htbp]
\begin{center}
\includegraphics[width=0.48\textwidth]{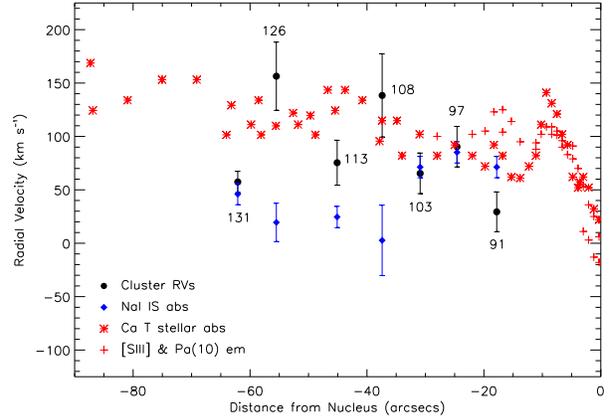}
\caption{Correspondence of star cluster velocities with the rotation curve for M82, as derived from 
infrared observations by \citet{mckeith93}. The cluster velocities appear to be consistent with the flat component 
of the curve. We also plot the velocities of the gas in the 
line of sight of the clusters. The distance in velocity-space between the clusters and the line of sight gas indicates the amount of dust and gas between the observer and the cluster. Therefore, this distance may be considered to represent the depth at which the cluster is found, with respect to the galaxy disk surface.
}
\label{plot:rot-curve}
\end{center}
\end{figure}

It has been claimed in the past that region B 
constitutes a bound region, that moves independently to the rest of the galaxy disk \citep{RdG01,RdG03a}. In order to gain 
some insight into this claim, we measure the 
radial velocities of the seven clusters for which 
spectroscopy is available, which we then 
proceed to plot over the rotation curve of the galaxy 
(see Figure.~\ref{plot:rot-curve}). This enables us to determine 
whether the clusters move in regular orbits. 
If the region was bound, the clusters would 
move in unison (solid body rotation); assuming a constant rotational 
velocity,  the outer boundary of the region would need 
to move twice as fast as the inner 
boundary for the clusters to move together. 

However, we find region B clusters to lie on the flat 
part of the curve, indicating that they move 
 along with the 
bulk of the galaxy. This means that the region 
cannot be self-bound, or kinematically independent to the nuclear gravitational well. 

\begin{figure}[bthp]
\begin{center}
\includegraphics[width=0.45\textwidth]{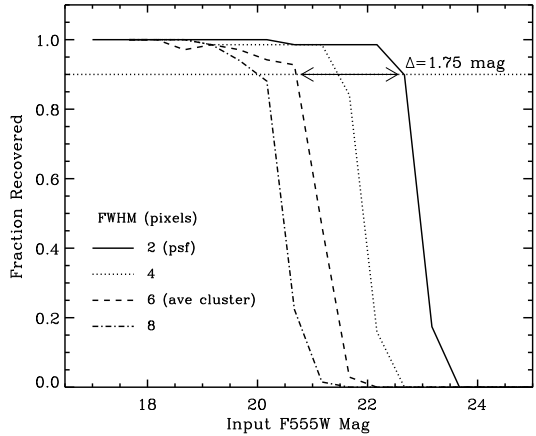}
\caption{This plot shows the fraction of input flux recovered from artificial sources of resolved (dotted, dashed and dashed-dotted lines) and unresolved (solid line) nature. The difference in recovered fraction between the average cluster and a point source (6 and 2 pixel FWHM respectively) is 1.75$^m$; \citet{RdG03a} treated M82 clusters as point sources, however our tests show \citep[as will be presented in][]{isk08b} that at the distance of M82 (3.6~Mpc), cluster surface brightness profiles are resolved. This discrepancy leads to a miscalculation of the completeness limit.  Correcting for this, we find no evidence for a turnover in the luminosity/mass distribution.}
\label{plot:det-limit}
\end{center}
\end{figure}

\section{Conclusions}
We have performed a detailed study of the stellar cluster population of M82 region~B, using spectroscopy and multi-band photometry, and conclude the following:

\begin{itemize}
\item The clusters are considerably younger than 
previously reported \citep{RdG03a}
, with ages in the range 10-300~Myr, 
peaking at 150~Myr. This is in agreement with 
the 220~Myr timescale for the last encounter 
with M81, the event that triggered the starburst.

\item We find significantly higher extinctions than previous 
studies, ranging up to $A_V\sim2.5$~mag. 
However, the region has a lower overall extinction 
compared to the rest of the disk, hence allowing 
a view deep into the body of the galaxy. 

\item The radial velocities of the clusters follow the 
galactic rotation curve, indicating that region B cannot 
be kinematically distinct from the rest of the 
disk.

\item Consideration of the detection limit for 
resolved clusters shows that the reported 
turnover in the mass/luminosity distribution of the 
clusters is caused by incompleteness effects (discussed in Figure.~\ref{plot:det-limit}). 

\end{itemize}

Overall, our findings contradict previous 
claims that present region B as having 
formed $\sim1$~Gyr ago, during an off-set, independent starburst 
episode. Our data strongly suggest that 
region B simply represents a line of sight into 
the galaxy, allowing a clear view of the cluster 
population of the disk, which undoubtedly formed during a 
galaxy-wide starburst.

The results presented here form part of a large spectroscopic study of the star cluster population of M82, which will be presented in \citet[][spectroscopy of 61 clusters]{isk08b}.

\bibliography{references}
\bibliographystyle{Spr-mp-nameyear-cnd}

\end{document}